\documentclass{elsarticle}

\usepackage{hyperref}
\usepackage{isotope}
\usepackage{graphicx}
\usepackage{subcaption}
\journal{Nuclear Instruments and Methods in Physics Research Section A}

\begin{document}

\begin{frontmatter}

\title{The low temperature performance of CsI(Na) crystals for WIPMs direct searches\tnoteref{t1}}
\tnotetext[t1]{Project study on a detector based on dual-light emitting crystal CsI(Na) and active shielding with liquid xenon for WIMPs direct searches (11305194) and Project nuclear recoil scale experiment on crystals for dark matter searches (11375220) supported by National Natural Science Foundation of China.This work is also supported in part by the CAS Center for
Excellence in Particle Physics (CCEPP).}

\author[IHEPaddress]{Xuan Zhang}

\author[IHEPaddress]{Xilei Sun
\corref{mycorrespondingauthor}}
\cortext[mycorrespondingauthor]{Corresponding author. Tel.: þ86 010 8823 5846; fax: þ86 010 8823 6423.
E-mail address: sunxl@ihep.ac.cn (X. Sun)}
\author[IHEPaddress]{Junguang Lu}
\author[IHEPaddress]{Pin Lv}

\address[IHEPaddress]{State Key Laboratory of Particle Detection and Electronics, Institute of High Energy Physics, CAS, Beijing 100049, China}

\begin{abstract}
Previous studies showed that CsI(Na) crystals have significantly different waveforms between alpha and gamma scintillations. In this work, the light yield and PSD capability of CsI(Na) scintillators as a function of the temperature down to 80 K has been studied. As temperature drops, the fast component rises and the slow component decreases. By cooling the CsI(Na) crystals, the light yield of high ionization events are enhanced significantly, while the light yield of background gamma events are suppressed. At 110 K, CsI(Na) crystal achieves the optimal balance between low threshold and good background rejection performance. The different responses of CsI(Na) to gamma and alpha at different temperatures are explained with self-trapped and activator luminescence centers. 
\end{abstract}

%CsI(Na); Crystal; Dark matter; Nuclear recoil; Particle discrimination; Self-trapped exciton

\end{frontmatter}

%%%%%%%%%%%%%%%%%%%%%%%%%%%%%%%%%%%%%%%%%%%%%%%%%%%%%%%%
%\linenumbers

\section{Introduction}
It is generally accepted that dark matter makes up about 27\% of the universe\cite{komatsu2011seven}. If dark matter consists of Weakly Interacting Massive Particles(WIMPs), they could scatter on the nuclei in various target material. The nuclear recoil energy (1-100keV) will be deposited in the target material. The energy threshold of detector material should be low enough to respond to such low nuclear energy.

Since such nuclear recoil event is extremely rare (estimated to be under 0.1 events/kg/day), it is vital to build shields with very low background materials outside the detector. However, there will still be inner background gamma events from the shielding material or the detector material itself inevitably. Thus, a background rejection power better than 10e-5 of the detector will be critical to distinguish nuclear recoils from the background electron recoils.

Cesium iodide is an extensively studied and used crystal scintillator. Doped CsI such as CsI(Tl) and CsI(Na) exhibit high light yield and good PSD capability. Previous studies showed that CsI(Na) exhibits very different responses to nuclear recoils and electron recoils [2]. The fast light of CsI(Na) from self-trapped exciton (STE) luminescence can be significantly excited by particles with a high ionization density, like nuclear recoil events. While the slow light of CsI(Na) can be attributed to the activator luminescence (Na emission), which are mainly excited by particles with a low ionization density, like electron recoil events. At room temperature, the decay time of fast light of CsI(Na) is about 20ns while that of the slow light is 670ns. This significant difference between nuclear and electron recoil pulse shapes makes CsI(Na) a possible candidate for dark matter searches. 

Since nuclear recoils are dominated by STE emission, the detection threshold for nuclear recoil events depends on the light yield of fast light of CsI(Na). It is possible to enhance the STE emission by lowering the temperature of the crystal. The scintillation light yield of pure CsI crystal is under 5000 photons/MeV at room temperature and reaches 100000 photons/MeV at 77K\cite{moszynski2003application,gridin2014channels}.

The aim of the present work is to study influence of low temperature (80 K to room temperature) on the light yield and PSD capability of CsI(Na) and find the optimum operating temperature of CsI(Na) for purpose such as the direct detection of WIMPs. Additionally, pure CsI crystal are tested as comparison.

%%%%%%%%%%%%%%%%%%%%%%%%%%%%%%%%%%%%%%%%%%%%%%%%%%%%%%%%
\section{Experimental}

In this study, the response of CsI(Na) and pure CsI crystals to gamma and alpha particles under low temperatures were tested. The emission was excited by Gamma-rays of 661.7 keV from a 0.5 micro-Curie \isotope[137]{Cs} source and alpha-rays of 5244 keV from a 5 micro-Curie \isotope[239]{Pu} source. The samples were placed into a home-made cryostat cooled with liquid nitrogen.

\subsection{Experimental set-up}
\begin{figure}
  \centering
    \includegraphics[width=4in]{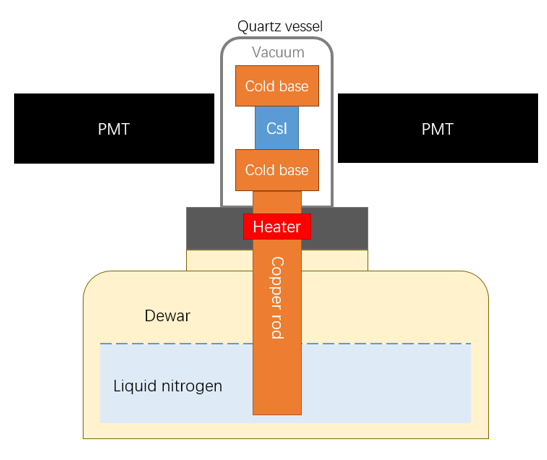}
  \caption{Experimental setup}
  \label{fig:ExperimentalSetup}
\end{figure}

Our setup is illustrated in figure \ref{fig:ExperimentalSetup}. A cubic shaped crystal sample was placed inside a cold copper base of the homemade cryostat. The dimensions of both CsI(Na) and pure CsI crystals were 25mm * 25mm *25mm. The doping concentration of Na in CsI(Na) was about 0.02\%. The left and right surfaces of crystal served as the light-emitting surfaces and the other 4 surfaces were cooled by the cold base. There was one layer of Enhanced Specular Reflector (ESR) film between the crystal and the cold base to minimize light loss during reflection inside the cubic crystal sample. The cold base was constantly cooled by a copper rod. The bottom end of the copper rod was soaked in liquid nitrogen in a dewar. A heater with PID power control was mounted in the joint part of the cold base and the copper rod. A Pt100 thermal sensor was attached to the bottom surface of the crystal cube to monitor the temperature and give feedback to temperature control system. The temperature control system can reach a precision of $\pm$0.1 K. The crystal and the cold base were encapsulated in a 2mm thick fused quartz vessel, which had an evenly distributed 80\% transmittance within a wide spectrum range. Since CsI(Na) are deliquescent crystals, the surfaces of the CsI(Na) sample were polished to eliminate the deliquescent outermost layer. The inside of fused quartz vessel was continually pumped to maintain the vacuum, which were vital to keeping low temperature as well as preventing further deliquescence of the crystal. Two 2-inch PMTs (R8778, Hamamatsu) were placed outside the quartz vessel facing the 2 light-emitting sides of the crystal to collect the scintillation light. The R8778 PMTs feature a high quantum efficiency of more than 30\% and an ultra-low background\cite{akerib2013ultra}. The quantum efficiency of R8778 PMT at 310 nm is close to its quantum efficiency at 410 nm.

\subsection{Data acquisition}
\begin{figure}
  \centering
    \includegraphics[width=4in]{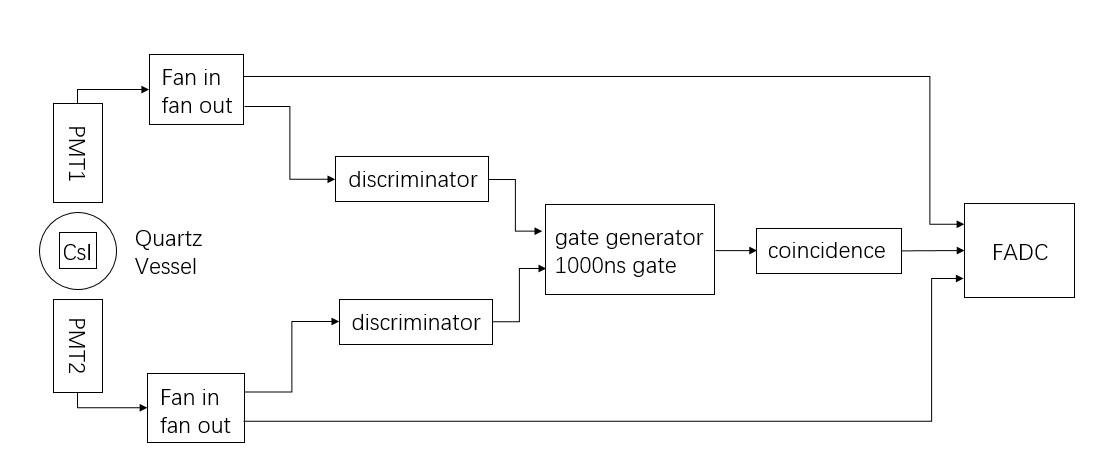}
  \caption{Data acquisition system}
  \label{fig:DataAcquisitionSystem}
\end{figure}

Figure \ref{fig:DataAcquisitionSystem} depicts the data acquisition system. The signals from the 2 PMTs were duplicated in a linear fan-in fan-out module (CAEN N625). One group of the 2 channel signals from the PMTs were fed to the waveform acquisition device. The other group of the signals were fed into a discriminator (CAEN N840) module and converted to logical pulse signals. The threshold of the discriminator was set to be 0.5 single photon voltage level. The 2 channel logical pulses from the discriminators were fed into a gate generator and generate 2 channel 1000 ns gate signals. The coincidence of the 2 gate signals triggered the acquisition of the 2 channel signals from the PMTs by the waveform acquisition device. The full waveforms were transfer to a PC and saved on hard disk. The waveform acquisition device in this study was either 4 synchronized digitizers (CAEN V1729a) or an oscilloscope (Tektronix DPO3054C) depending on the acquisition window width. Each digitizer had a memory depth of 1.26 us with a sampling frequency of 2 GS/s. The 4 synchronized digitizers can acquire 5 us long waveforms at 800 Hz. The oscilloscope can acquire 400 us long waveforms with a sampling frequency of 2.5 GS/s at less than 1 Hz.

%%%%%%%%%%%%%%%%%%%%%%%%%%%%%%%%%%%%%%%%%%%%%%%%%%%%%%%%
\section{Results}

The temperature of pure CsI and CsI(Na) samples were first brought down to 80 K and gradually rise to room temperature.

\subsection{Waveform of CsI(Na)}
\begin{figure}
    \centering
    \begin{subfigure}[b]{0.7\textwidth}
        \includegraphics[width=\textwidth]{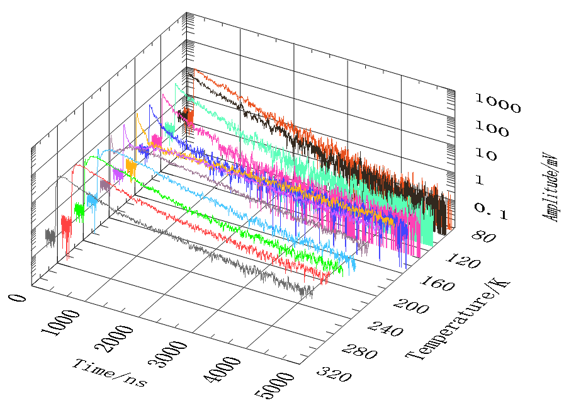}
        \caption{The waveforms of $\gamma$-scintillations from CsI(Na) at different temperatures. Each waveform is the averaged waveform of 50 events at the 661.7 keV peak.}
        \label{fig:WVFMCsINaGamma}
    \end{subfigure}
    ~ %add desired spacing between images, e. g. ~, \quad, \qquad, \hfill etc. 
    \begin{subfigure}[b]{0.7\textwidth}
        \includegraphics[width=\textwidth]{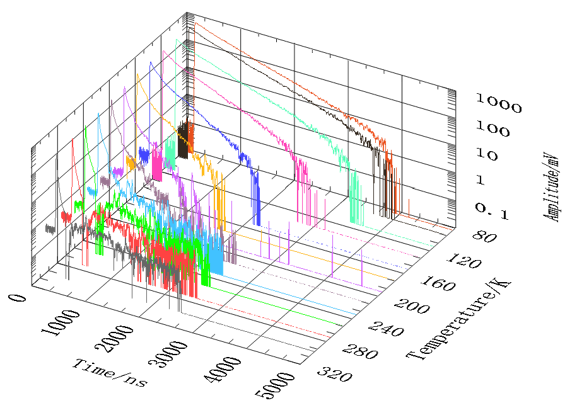}
        \caption{The waveforms of $\alpha$-scintillations from CsI(Na) at different temperatures. Each waveform is the averaged waveform of 50 events at the 5.2 MeV peak.}
        \label{fig:WVFMCsINaAlpha}
    \end{subfigure}
    \caption{Waveforms of CsI(Na)}\label{fig:WVFMCsINa}
\end{figure}

Figure \ref{fig:WVFMCsINa} shows the waveform plots of $\gamma$-scintillations (661.7 keV) and $\alpha$-scintillations (5.2 MeV) from the CsI(Na) crystal. 
The waveforms of $\alpha$-scintillations consist of a fast component and a slow component while of $\gamma$-scintillations. As the temperature decreases, the fast component of $\alpha$-scintillation increase significantly and both $\alpha$-scintillations and $\gamma$-scintillations slow down.

As the temperature decreases, both of the fast and slow component of α-scintillation slow down. The decay time of the fast component increased from 16 ns at 298 K to ~700 ns at 80 K. The profile of slow component becomes flat and is hard to be distinguished from the fast component under 190 K. 

For the $\gamma$-scintillations from the CsI(Na) crystal, the slow components dominate the waveforms at higher temperatures. As the temperature decreases, slow component becomes lower and flatter while fast component shows up. In Figure \ref{fig:WVFMCsINaGamma}, notable fast component can be seen on the leading edge of the $\gamma$-scintillations under 210 K. At 80 K, the decay time of fast component is 750 ns.

\subsection{Waveform of pure CsI}
\begin{figure}
    \centering
    \begin{subfigure}[b]{0.7\textwidth}
        \includegraphics[width=\textwidth]{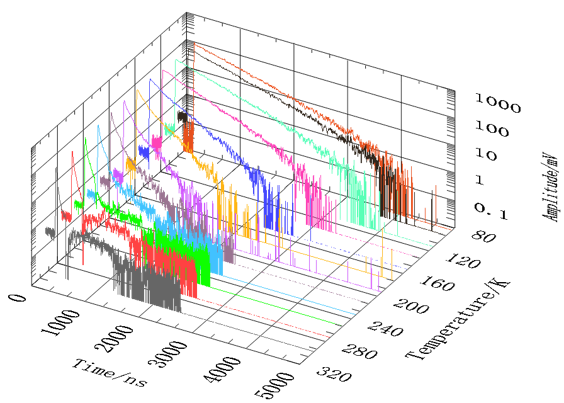}
        \caption{The waveforms of $\gamma$-scintillations from pure CsI at different temperatures. Each waveform is the averaged waveform of 50 events at the 661.7 keV peak.}
        \label{fig:WVFMpureCsIGamma}
    \end{subfigure}
    ~ %add desired spacing between images, e. g. ~, \quad, \qquad, \hfill etc. 
    \begin{subfigure}[b]{0.7\textwidth}
        \includegraphics[width=\textwidth]{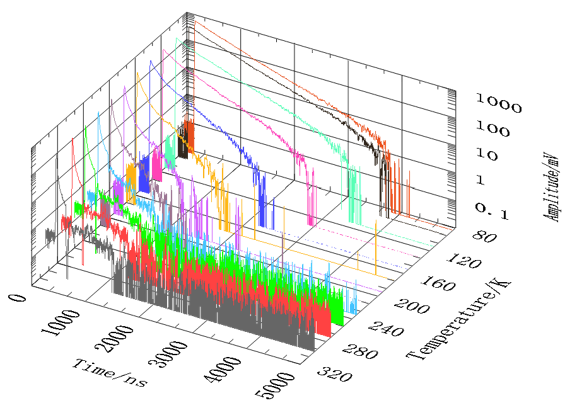}
        \caption{The waveforms of $\alpha$-scintillations from pure CsI at different temperatures. Each waveform is the averaged waveform of 50 events at the 5.2 MeV peak.}
        \label{fig:WVFMpureCsIAlpha}
    \end{subfigure}
    \caption{Waveforms of pure CsI}\label{fig:WVFMpureCsI}
\end{figure}

Figure \ref{fig:WVFMpureCsI} shows the waveform plots of $\gamma$-scintillations (661.7 keV) and $\alpha$-scintillations (5.2 MeV) from the pure CsI crystal. Under both circumstances, only fast components can be seen and the waveforms slow down and spread as the temperature decreases. At 80K, the decay time is 620 ns for $\alpha$-scintillations and 700 ns for $\gamma$-scintillations. The $\alpha$-scintillations and $\gamma$-scintillations from pure CsI crystal are very similar to the fast components of scintillations  from CsI(Na) crystal.

As a comparison, waveform of CsI(Na) has both fast and slow components while that of pure CsI only has fast component, suggesting that the slow component can be attributed to the Na activator. As a dopant, Na activators have a larger average spacing than the in CsI(Na). And the fact that gamma events yield significantly more slow component than alpha events is consistent with the discussion on the mechanism in our previous papers\cite{xi2011fast,sun2011neutron}.

\subsection{Light yields}

We can integrate the waveform over time from trigger point to 5 us after the trigger point as measure of the light yield of the fast component. And the integration with a 400 us integration range point can be viewed as a measure of the total light yield, which include both the fast component and the slow component.

\begin{figure}
  \centering
    \includegraphics[width=4in]{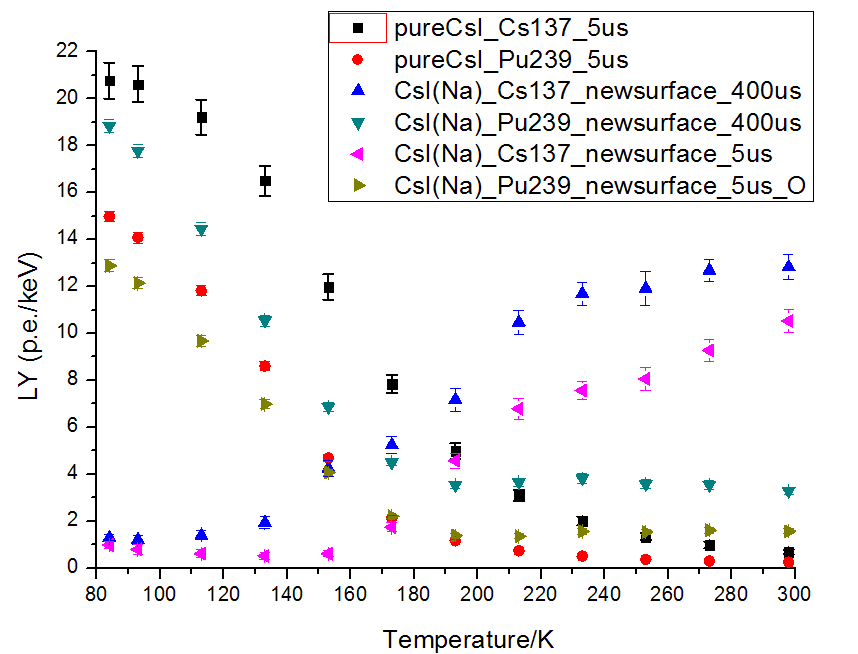}
  \caption{Light yield of pure CsI and CsI(Na) as a function of temperature. The value of light yield has been multiplied by a factor of 3.86, which is the difference between the light collection efficiency of the PMTs directly coupled to the crystals and the PMTs placed outside the quartz vessel. Black squares represent the light yield from the first 5 us integration of the $\gamma$-scintillations from pure CsI. Red dots represent the light yield from the first 5 us integration of the $\alpha$-scintillations from pure CsI. Blue and cyan triangles represent the light yield from the 400 us integration of scintillation waveforms from CsI(Na). Magenta and yellow triangles represent the light yield  from the first 5 us integration of the scintillation waveforms from CsI(Na).}
  \label{fig:lightYield}
\end{figure}

In figure \ref{fig:lightYield}, the light yield of CsI(Na) and undoped CsI crystal are plotted as a function of temperature. The first 5 us waveform of the scintillation is considered to be mainly dominated by the fast component. The whole 400 us waveform include both fast and slow components.
As temperature falls, the light yield of both fast and slow component of $\alpha$-scintillations from CsI(Na) crystal rises while the $\gamma$-scintillation one drops. From 298 K to 80 K, the light yield of $\alpha$-scintillations was increased by 8.25 and 5.56 times for the fast component window and full waveform respectively, while the light yield of $\gamma$-scintillation fell 9.2\% and 10\% respectively.
For pure CsI, the light yield of both $\alpha$-scintillations and $\gamma$-scintillations rises as temperature falls. At 80 K, the light yield of $\alpha$-scintillations and $\gamma$-scintillations  was increased by a factor of 62 and 30.4 respectively, compared to that at 298 K.

\subsection{Pulse shape discrimination}

\begin{figure}
  \centering
    \includegraphics[width=4in]{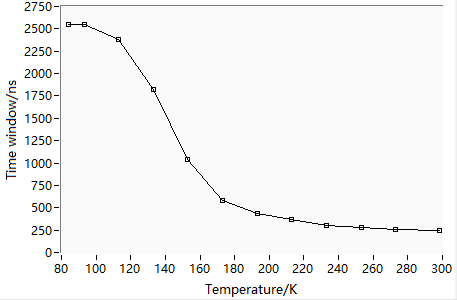}
  \caption{Optimized fast component window time at different temperatures. The window time is the time of intersection of the normalized waveforms of $\gamma$-scintillation and $\alpha$-scintillation from CsI(Na).}
  \label{fig:timeWindow}
\end{figure}

We use the ratio of the fast component to the total light yield of CsI(Na) to characterize the pulse shape discrimination factor. The decay time of fast component varies with the temperature. The integral window width for the fast component is scanned to achieve max difference of the fast/total ratio. The optimal time window is plotted as a function of temperature in figure \ref{fig:timeWindow}. As temperature falls, the time window curve rises. There is a rapid change between 120 K and 180 K. 
%In figure \ref{fig:WVFMpureCsI}, the waveform of pure CsI crystals 
% and falls before 210 K. This is due to the domination of slow %component above 210 K. The time window monotonically increases %below 210 K which is caused by the spread fast component.

\begin{figure}
  \centering
    \includegraphics[width=4in]{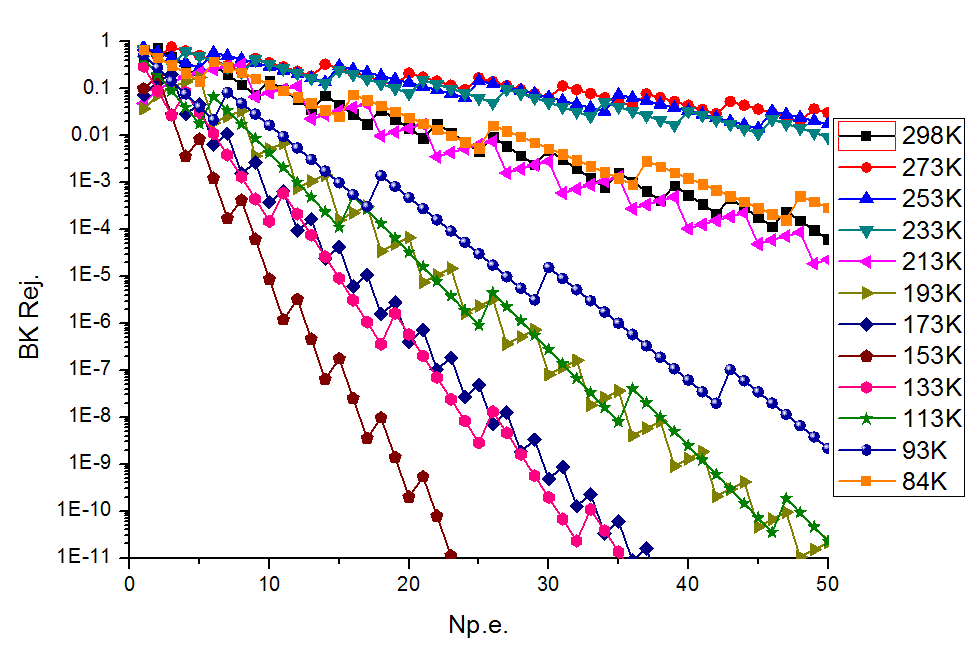}
  \caption{Background rejection ratio at different temperatures}
  \label{fig:rejectionRatio}
\end{figure}

With the time window width determined, the rejection ratio under different temperature are calculated with a minimum efficiency of 60\%. The results are showed in figure \ref{fig:rejectionRatio} as a function of sampling number of photo electrons. As can be seen in figure \ref{fig:rejectionRatio}, CsI(Na) crystal achieves best background rejection performance at 150 K, which can be low as $10^{-6}$ at 10 p.e. As shown in figure \ref{fig:lightYield}, with a further decrease in temperature, the light yield of fast component increases significantly. We need to weigh the benefit of lower threshold for nuclear recoil events with the cost of lower background rejection performance.  Considering the low energy of nuclear events, 110 K could be the optimal operating temperature for future CsI(Na) based dark matter detector.

\begin{figure}
  \centering
    \includegraphics[width=4in]{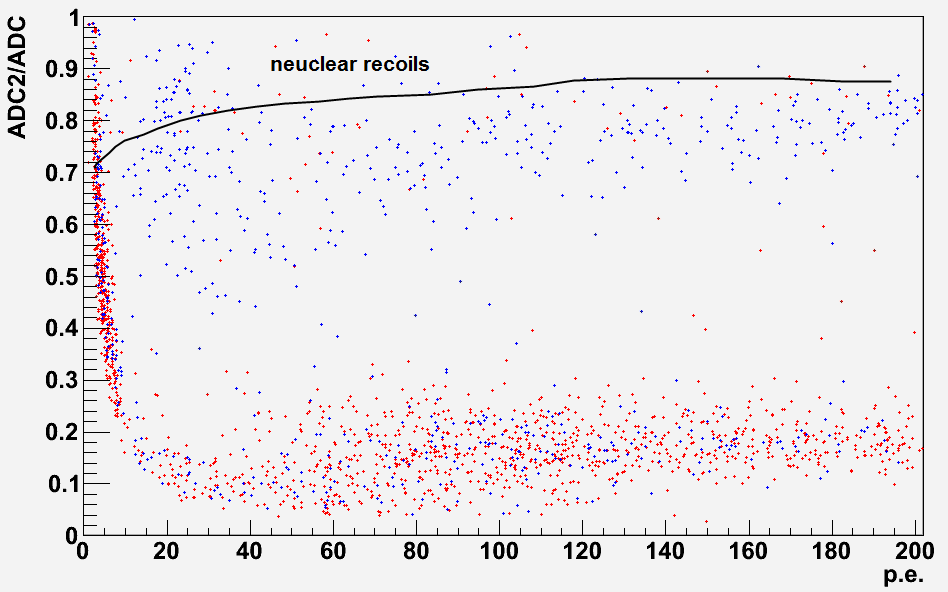}
  \caption{PSD scatter plot of gamma and alpha events at 160 K. Red dots are gamma events, and blue dots are alpha events}
  \label{fig:PSDscatter}
\end{figure}

Figure \ref{fig:PSDscatter} shows the PSD scatter plot at 160 K with a fast component integral window of 1.26 us. The $\alpha$-scintillations and $\gamma$-scintillations events form two separate belt zones. The electron recoil events bend up sharply below 10 p.e. and cross the nuclear recoil event zone, which is consistent with figure \ref{fig:rejectionRatio}. Moreover, there are odd but non-negligible number of electron events mixed in the nuclear recoil event zone.

%%%%%%%%%%%%%%%%%%%%%%%%%%%%%%%%%%%%%%%%%%%%%%%%%%%%%%%%
\section{Discussion}

According to Payne\cite{payne2009nonproportionality} and Gridin\cite{gridin2014excitonic}, after the creation of electron-hole pairs, there are many possible branches that the created electrons and holes can interact with each other and with the activators. The electrons and holes can form excitons. The holes can be self-trapped recombine with an electron and form self-trapped exciton (STE). Activators can capture electrons and holes and jump to excited states.

In pure CsI, STE is the dominated luminescence center. If the distance between an electron and a hole is $r$, the electron will diffuse under the Coulomb field and recombine with the hole with the probability  $p=1-e^{(-R_{Ons}/r)}$ $(R_{Ons}=e^2/(4πεε_0 k_B T)$, $R_{Ons}$  is the Onsager radius, where $e$ is the electron charge, $\varepsilon$ is the static dielectric permeability, $k_B$ is the Boltzmann constant and $T$ is the temperature). As the temperature falls, the Onsager radius $R_{Ons}$ rises, electrons at a longer distance from the hole can diffuse to recombine with the holes. This means more electrons and hole can form excitons and enhances the STE emission and fast components as temperature drops.

In CsI(Na), both STE and activator (Na) luminescence can take place. In order to form the excited state of activator, the activator will have to capture an electron and a hole, or a hole and an electron sequentially, which limits the rate of production of excited states of activator. The slow component is generated by activator luminescence. Since the total number of electrons and holes is fixed, the STE and activator luminescence are two competing processes. The prevailing process depends on the concentration of activator ($n_A$), electrons ($n_e$) and holes ($n_h$). The electrons and holes are formed in pairs, so $n_e=n_h$. For particles with high ionization density, $n_h\gg n_A$, the electrons are most likely to be captured by self-trapped holes, so the STE luminescence dominates. That explains why the waveform of $\alpha$-scintillation is dominated by fast component and very similar to waveforms from CsI. 

For particles with low ionization density like gamma or electrons, $n_h\ll n_A$, the electrons have a much higher chance to be captured by activators than recombining with self-trapped holes and form STE. So the slow component prevails in the waveform of $\gamma$-scintillation from CsI(Na).

The capture radius of activator changes little as temperature drops, because it is caused by the dipole polarization of the neutral activator center in the electron Coulomb field\cite{payne2009nonproportionality}. As the temperature falls, and Onsager radius rises but the activator capture radius changes little. So the STE emission rises and the activator emission drops. 

By cooling the CsI(Na) crystals, we can enhance the quenching factor for high ionization density events, while suppressing the quenching factor for gamma background. The significant increase in light yield of $\alpha$-scintillations from CsI(Na) at low temperatures lower the energy threshold to detect nuclear recoil events. 

%On the other hand, the waveforms slow down and spread to discrete single photons at low temperatures. To avoid events overlay and piled up, passive background shields will be a must for large CsI(Na) detectors working at low temperature. As the temperature falls, the trailing waveforms turn into series of discrete single photos. So a data acquisition system features high sample rate (1 GSample/s), very wide acquisition time window (100 us) and low noise would be necessary to build a dark matter detector based on CsI(Na).

%%%%%%%%%%%%%%%%%%%%%%%%%%%%%%%%%%%%%%%%%%%%%%%%%%%%%%%%
\section{Conclusions}

CsI(Na) crystals exhibit discriminability between alpha and gamma scintillation at temperatures as low as 80 K. As temperature goes down, the light yield of alpha scintillation increases but that of gamma scintillation decreases. The response of CsI(Na) and pure CsI to gamma and alpha at different temperatures can be explained with the competing processes of forming STE and activator luminescence centers. By cooling the CsI(Na) crystals, light yield of high ionization density or nuclear recoils events is enhanced, while light yield of the gamma background events is suppressed. PSD capability of CsI(Na) reaches the peak at 110 K, which can be the optimum operating temperature for future CsI(Na) based dark matter detector.

%%%%%%%%%%%%%%%%%%%%%%%%%%%%%%%%%%%%%%%%%%%%%%%%%%%%%%%%
\section*{References}

%%%%%%%%%%%%%%%%%%%%%%%%%%%%%%%%%%%%%%%%%%%%%%%%%%%%%%%%
\bibliography{mybibfile}

\end{document}